\documentclass[12pt]{article}
\usepackage{graphicx}
\begin{document}
\title{On UHECR energy estimation algorithms based on the measurement of electromagnetic component parameters in EAS}
\author{A.A. Ivanov\footnote{e-mail: ivanov@ikfia.ysn.ru}\\
\\
\parbox{25pc}{\small
\centering
\mbox{Yu.G. Shafer Institute for Cosmophysical Research and Aeronomy}\\ Yakutsk 677980, Russia\\}}
\date{}
\maketitle
\begin{abstract}
Model calculations are performed of extensive air shower (EAS) component energies using a variety of hadronic interaction parameters. A conversion factor from electromagnetic component energy to the energy of ultra-high energy cosmic rays (UHECRs) and its model and primary mass dependence is studied. It is shown that model dependence of the factor minimizes under the necessary condition of the same maximum position and muon content of simulated showers.
\end{abstract}

\section{Introduction}
UHECR particles hitting Earth atmosphere produce a cascade of secondary particles, the small part of which is detected on the ground with EAS array. The energy of the primary particle, $E_0$, is distributed among the shower components. The most of the energy deposit is due to ionization and excitation of the air molecules caused by electromagnetic component.

The primary energy estimation algorithms are based mainly on the measurement of the shower parameters related to electromagnetic component energy, $E_{em}$. Such an approach is realized in High Resolution Fly's Eye (HiRes)~\cite{Sokolsky}, Pierre Auger Observatory (PAO)~\cite{Auger} and the Yakutsk array~\cite{Mono,CRIS} experiments. Future applications are planned in the Telescope Array and satellite projects.

In all these experiments a fraction of the primary energy cannot be measured because it is carried away by hadrons, muons etc., unobservable with array detectors. This 'missing energy' and conversion factor $E_{em}/E_0$ can be calculated modeling a cascade in atmosphere, as was done in~\cite{CRIS,Risse,Barbosa,Muniz}. However, only QGSJET and SIBYLL models have been applied in these simulations, setting aside the model dependence of the conversion factor.

In the present work the energies of air shower components are calculated using extremely different hadronic interaction models, but having the same maximum position and muon content of the resultant showers. It is shown that the model dependence of the ratio $E_{em}/E_0$ minimizes in this case. The variability of the ratio is estimated due to models and the primary nucleus mass. The experimental uncertainties in the electromagnetic component energy measurements are discussed. In particular, the energy spectra measured with the Yakutsk and HiRes arrays are shown to be coincident within experimental errors.

\section{Basic experimental data}
Relativistic electrons of the shower induce Vavilov-Cherenkov radiation (Cherenkov light) in the atmosphere. The total flux of light, $Q_{tot}$, can be used as an estimator of the electromagnetic component energy in EAS. This method realized in the Yakutsk array experiment is based on the relation between $Q_{tot}$ and $E_{em}$ largely independent of the model~\cite{Mono,CERN}. The energy fraction dissipated in the ground is estimated using the tail of cascade curve measurement in inclined showers.

Another method, namely fluorescence technique, is applied in HiRes and PAO/FD experiments where the fluorescence light emitted by wounded nitrogen molecules along the trajectory of the shower is collected by mirrors and received by photomultiplier pixels. The light intensity is proportional to the number of shower electrons, so the cascade curve, $N_e(t)$, is scanned by fired pixels and ionization in the atmosphere (ionization integral) can be estimated:
$$
E_i^{t_0}=\int_0^{t_0}N_e(t)dt,
$$
where $t$ is slant depth along the shower axis in units of radiation length (36.7 g/cm$^2$~\cite{PDG}); $t_0$ is the thickness of atmosphere; energies are in units of critical energy ($\epsilon_0=86$ MeV). As it was shown in cascade theory~\cite{Belenky}, the electromagnetic component energy is equal to ionization integral $E_{em}=E_i^\infty$; that is why the energy $E_{em}$ is actually evaluated in these experiments.

\begin{figure}[tb]
\begin{center}
\includegraphics[width=0.6\columnwidth]{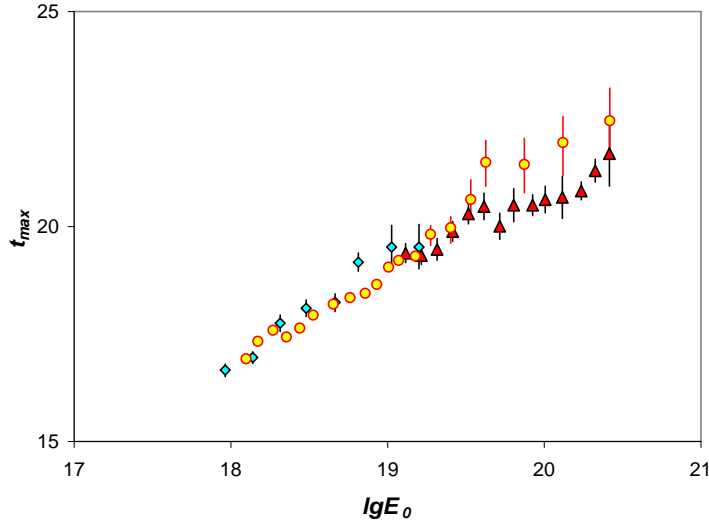}
\end{center}
\caption{\footnotesize Shower maximum depth in the atmosphere as a function of EAS primary particle energy.
HiRes~\cite{Sokolsky} (triangles) and HiRes/MIA~\cite{CascadeCurve} (rhombuses)
results are shown in comparison with the Yakutsk array data (circles).}
\label{fig:Tmax}
\end{figure}

Measurements of the shower maximum position in the atmosphere as a function of the primary particle energy, $t_{max}(E_0)$, and the number of muons at observation level (usually as a ratio to the number of electrons, $N_\mu/N_e$) are the basic parameters to be guided by performing the cascade simulations in the UHE region. The only data available in the region are those given by the HiRes, AGASA and Yakutsk experiments~\cite{Sokolsky,Mono,Nagano}. In the HiRes case the position of the shower maximum is measured by stereo system of two Eyes. On the contrary, the Yakutsk array detectors are measuring the Cherenkov light, and the lateral distribution form of the light spot on the ground is used to derive $t_{max}$ in this case (Fig.~\ref{fig:Tmax}).

\begin{figure}[tb]
\begin{center}
\includegraphics[width=0.6\columnwidth]{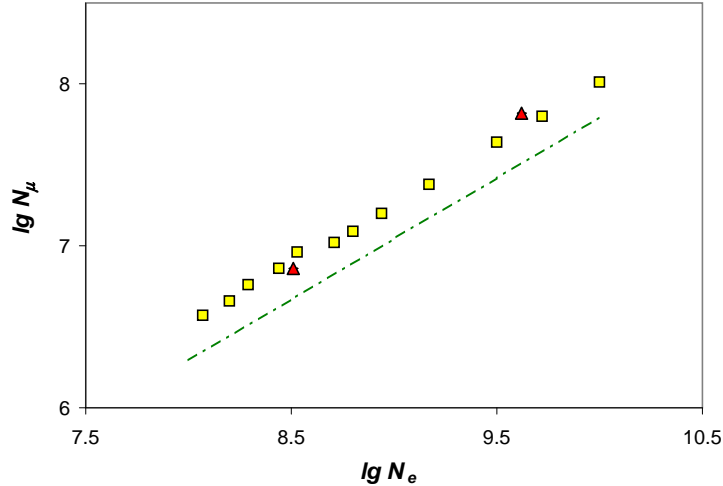}
\end{center}
\caption{\footnotesize The number of muons ($E_\mu>1$ GeV) vs the number of electrons
at the depth $t_0=27.8$. Current and previous~\cite{Mono} Yakutsk array
data are shown by squares and triangles. AGASA data
approximation~\cite{Nagano} is given by the dash-and-dot line.}
\label{fig:NmNs}\end{figure}

The number of electrons and muons at the ground level is measured by scintillation detectors/proportional counters (shielded with Fe/concrete and ground to detect muons) of the Yakutsk and AGASA arrays (Fig.~\ref{fig:NmNs}).

Experimental errors in $t_{max}$ measurement are $0.82$ to $1.38$ in the energy range $E_0\epsilon_0\in(10^{17},10^{19})$ eV; $N_\mu/N_e$ ratio differs up to $\sim2$ times between Yakutsk and AGASA data while the difference in $N_\mu$ itself is around 25\%~\cite{Nagano}. The main source of the discrepancy is the total number of charged particles on the ground which is estimated using the particle density measured at the shower periphery in both experiments.

\section{Modeling the cascade in atmosphere}
As was shown in previous calculation of the energy balance of EAS components~\cite{CRIS}, the dependence of primary energy deposited in the atmosphere on hadronic interaction models (i.e. multiplicity of secondaries, cross sections, etc.) can be parameterized via the longitudinal shower development characteristics $t_{max}$ and $N_\mu/N_e$. This is a consequence of the close connection between  $E_{em}$ and the ionization integral. In this case (as far as other shower parameters are not concerned), we can go far beyond the bounds of conventional models.

Namely, we can use in very high energy region a variety of arbitrary extrapolations of accelerator data in order to apply hadronic interaction models with only restriction of the same resultant shower maximum and the muon content. It will allow to estimate the real limits of the ratio $E_{em}/E_0$ variations using distinctly different models in addition to 'standard' models fitted to describe the same experimental data.

In this work three hadronic interaction models which are characterized by extremely different forms of the rapidity, $y$, distribution of the 'sea' secondaries in pionization region are used: i) Gauss model with normal rapidity distribution; ii) Delta model with equal rapidities of $n_s$ secondaries and iii) Flat distribution model with $f(y)=const$. All other parameters of the models are flexible in order to get $t_{max}$ and $N_\mu/N_e$ of the shower fitting experimental data within errors. In the Flat distribution model, for instance, the pionization region width and the fraction of secondary pions, kaons and nucleons in multiple production processes have been adjusted in addition to cross sections and fragmentation coefficients to get the output EAS observables desired.

Numerical solution of hadron transport equations is obtained using the resolvent of Volterra equation on the rectangular lattice $\{t_i,y_k\}$~\cite{Resolvent}. To calculate the number of electrons in the shower at a depth $t$, Greisen's formula is employed. Electromagnetic component energy is assumed to be equal to the energy of neutral pions, the decay of which into gamma-quanta initiates the component. The muon and neutrino energy is calculated via the energy fraction taken from the decay of charged pions and kaons. The hadronic component energy is summed up at each layer $t_i$.

It is well-known that $t_{max}$ is strongly fluctuating parameter of the shower. Monte Carlo simulations using CORSIKA code give RMS deviation $1.9$ and $1.1$ for the proton and Fe initiated showers of the same $E_0$, correspondingly~\cite{Ostap}. On the contrary, the primary energy fraction deposited in atmosphere is fluctuating much lesser because it is related to the integral along the shower trajectory.

\begin{figure}[t]
\begin{center}
\includegraphics[width=0.6\columnwidth]{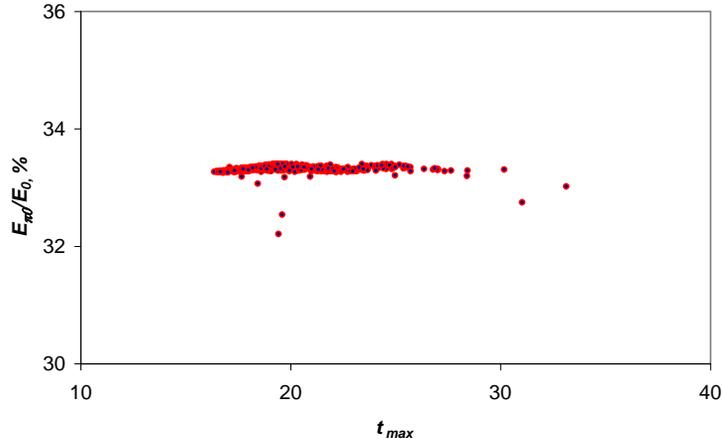}
\end{center}
\caption{\footnotesize Event-by-event fluctuations in a leading fragment approximation.}
\label{fig:Correlation}
\end{figure}

In order to estimate the influence of cascade fluctuations on the ionization integral, a leading fragment approximation in the simplest case of the primary nucleon is used - only fluctuations in inelasticity and interaction points of the EAS primary particle are considered, along suggestions given in~\cite{Dedenko}. Bearing in mind the isotopic invariance of multiple production processes and multiplicity distribution, we derive the ionization integral event-by-event distribution, $f(E_{em})$ with fixed $E_0$, as a ($n=x_0/\lambda\sim10$ - fold) convolution of inelasticity fluctuations. Resultant distribution is a Gaussian with average energy fraction
$$\frac{E_{\pi^0}}{E_0}=\frac{1-\bar{K}^n}
{1-\bar{K}}\overline{K_{\pi^0}},$$
where $K$ is an elasticity in nucleon interactions; $K_{\pi^0}$ is a projectile energy fraction carried out by neutral secondary pions. The RMS deviation is $\approx0.1\%$ as illustrated in Fig.~\ref{fig:Correlation}. We neglect fluctuations in the rest of $E_0$ distributed among the shower components because of averaging over the huge number of shower particles.

\section{Results}
\begin{figure}[t]
\center{\includegraphics[width=0.6\columnwidth]{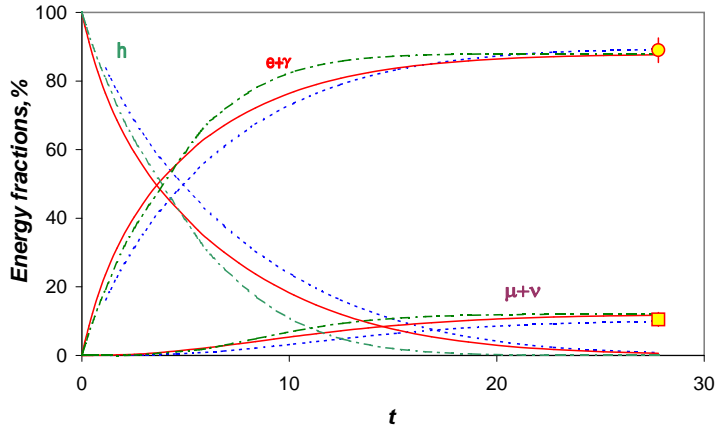}}
\caption{\footnotesize Energy fractions of EAS components as a function of the depth in the atmosphere.
Model calculation results are given for hadronic ($h$), electromagnetic ($e+\gamma$)
components and muons and neutrinos ($\mu+\nu$). The solid line refers to Gauss model,
dotted line to Delta model and dash-and-dot line to the Flat rapidity distribution model.
Experimental data of the Yakutsk array are shown at $t=27.8$: ionization in the shower (circle) and the energy of muons+neutrinos (square)~\cite{Eem}.}
\label{fig:Balance}
\end{figure}

We are interested in the ionization integral and the energy deposited to muons+neutrinos in a shower. These values calculated for our models ($E_0\epsilon_c=10^{18}$ eV) are presented in Fig.~\ref{fig:Balance}. The hadronic component energy is less than $0.01E_0$ at sea level.

Ionization in the atmosphere, $E_i^{t_0}$, is closely related to the total air Cherenkov light flux on the ground~\cite{Mono,CRIS} and is measured with Cherenkov light detectors of the Yakutsk array. A point in Fig.~\ref{fig:Balance} is given with experimental error in the aggregate of calibration, atmospheric extinction of light uncertainty, etc.

The energy of muons and neutrinos is estimated as a sum of $E_\mu$ measured on the ground and model calculation results giving the energy of neutrinos, ionization and decay of muons~\cite{Erlykin}. Absolute experimental $E_{\mu+\nu}$ uncertainty ($\sim2\%$) is based on the difference between Yakutsk and AGASA $N_\mu$ measurements.

\begin{figure}[t]
\center{\includegraphics[width=0.6\columnwidth]{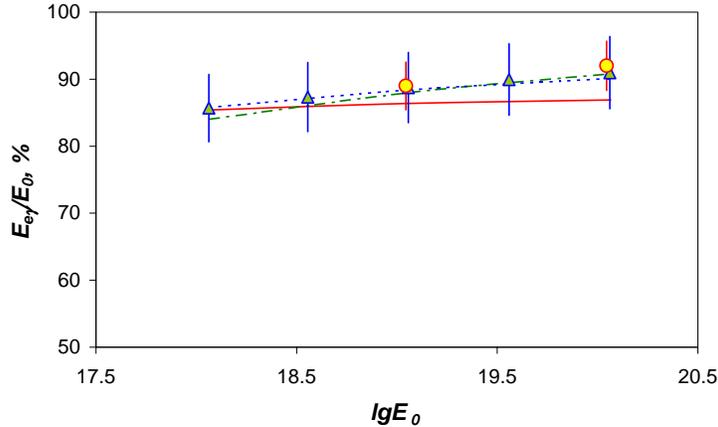}}
\caption{\footnotesize Energy fraction of electromagnetic component vs the primary particle energy.
The hadronic interaction model results are shown by the same signs as in Fig.~\ref{fig:Balance}. A HiRes estimation is shown by triangles~\cite{Sokolsky}. Simulation uncertainty 6\% due to unknown primary nucleus mass is assigned to points. Two circles at $E_0\epsilon_0=10^{18}$ and $10^{19}$ eV illustrate $E_{em}/E_0$ evaluation based on the Yakutsk array data~\cite{CRIS}. Experimental errors are shown by vertical bars.}
\label{fig:EgE0}
\end{figure}

Another presentation of the results is given in Fig.~\ref{fig:EgE0}. Electromagnetic component energy is $E_{em}=E_i^{t_0}+E_l$, where $E_l$ is the energy of electrons and photons dissipated in the ground. To estimate it we need the total number of electrons at $t_0$ and its attenuation length: $E_l\simeq\lambda_{N_e}N_e(t_0)$. The result based on the Yakutsk array data is shown in Fig.~\ref{fig:EgE0} by circles.

The Yakutsk array data are used to estimate the energy of muonic component. Energy remainder deposited to undetectable hadrons and neutrinos is estimated using the cascade modeling. In the HiRes case, the energy fraction $E_{em}/E_0$ is exclusively the product of model simulations because the only component measured is the fluorescence light. Since CORSIKA/QGSJET code is used to estimate the conversion factor to primary energy, triangles represent QGSJET model results. EAS simulation with SIBYLL model gives the ratio with difference less than $1.6\%$~\cite{Barbosa}.

The main feature of results given in Figs.~\ref{fig:Balance},\ref{fig:EgE0} is that different models lead to approximate fractions of the primary energy assigned to electromagnetic component, if $t_{max}$ and $N_\mu/N_e$ are coincident. The model uncertainty is less than 3\% in the interval $E_0\epsilon_0\in(10^{17},10^{19})$ eV where the experimental data are available; the energy dependence of $E_{em}/E_0$ is limited to $\sim2\pm1\%$ per decade, while a spread at $10^{18}$ eV is within $87\pm1\%$.

Additionally, the ionization integral and $E_{\mu+\nu}$ are varying with the primary particle mass, in spite of the same $t_{max}$ for different nuclei. We estimated the variance to be less than $6\%$ if the primary mass is $1\leq A\leq56$. The sum of two uncertainties (from the model and primary mass) is $\delta E_{em}/E_0\leq7\%$. We have not considered here showers initiated by other primaries, such as photons, neutrinos or mini black holes, because their muon content is a subject of the alternate investigation.

Due to the negligible fraction of hadronic component energy at sea level, the remainder of $E_0$ is transferred to muons and neutrinos. As a consequence, $E_{\mu+\nu}/E_0$ is nearly model independent, too, but the muon component energy measurable on the ground is more variable in different models.

\section{Experimental uncertainties}
The experimental errors in the shower maximum detection and the number of muons at $t_0$ lead to the uncertainty of calculated ratio $E_{em}/E_0$ (we will pass over $\delta E_{\mu+\nu}/E_0$ considering it congruous). We have modeled this varying the multiplicity of secondaries and cross sections which result in the changes of $t_{max}$ and $N_\mu(E>1 GeV)$ comparable to experimental uncertainties. Resultant $\delta E_{em}/E_0$ turned out to be below $5\%$ in the case of depth variation $\delta t_{max}\leq1.36$, and $3\%$ for $\delta N_\mu/N_\mu\leq0.25$. So we have assumed the aggregate experimental uncertainty due to $t_{max}$ and $N_\mu$ measurement errors to be below $6\%$.

The greatest uncertainty source is the ionization integral measurement itself. For the Yakutsk array data it comprises of errors due to Cherenkov light measurement and the number of charged particles on the ground~\cite{CRIS,CERN}: uncertainty in atmospheric transparency (15\%); detector calibration (21\%) and total light flux measurement (15\%) errors; an uncertainty in the number of electrons reaches $60\%$ because the only measurable parameter is the particle density beyond hundred meters from the shower core. Resultant $E_{em}/E_0$ estimation uncertainty ($\sim30\%$) is the sum of all these errors weighed with the shower component energies.

The HiRes group claims a systematic uncertainty $\sim20\%$~\cite{Sokolsky} aggregated of errors in the absolute calibration of the photo-tubes ($10\%$), the yield of the fluorescence process ($10\%$), modeling of the atmosphere, and so on.

\begin{figure}[t]
\center{\includegraphics[width=0.6\columnwidth]{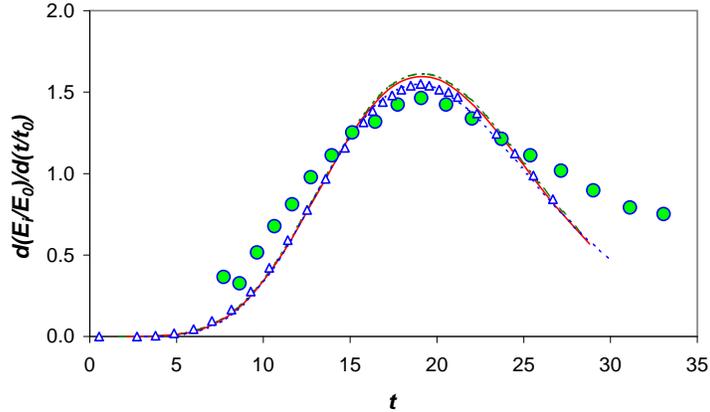}}
\caption{\footnotesize The average longitudinal profile of EAS (circles,$E_0\epsilon_0\in(10^{17},10^{18})$ eV) measured in the HiRes/MIA hybrid experiment~\cite{CascadeCurve} and shower cascade curves calculated in models. Signs for models are the same as in previous Figures. The average cascade curve of electrons (triangles) for proton-initiated vertical showers of energy $E_0\epsilon_0=10^{17.5}$ eV simulated using CORSIKA/QGSJET~\cite{Pierog} is given for comparison. Curves are adjusted to $t_{max}=19.1$.}
\label{fig:Cascade}
\end{figure}

However, a comparison of model predictions with the HiRes/MIA measurement of the average cascade curve in the interval $(10^{17},10^{18})$ eV (Fig.~\ref{fig:Cascade}) reveals a discrepancy between calculated and measured curves. Substantially, a contradiction in the cascade curve can be resolved using the attenuation length of electrons measured at sea level. Equal intensity cuts method gives the value $\lambda_{N_e}=5.6\pm0.9$ in the interval $t\in(27.25,29.97)$ for the Yakutsk array data~\cite{Mono}, while the HiRes curve has $\lambda_{N_e}=15.9\pm1.1$. Models predict the length $\lambda_{N_e}\in(4.7,5.4)$ in agreement with the Yakutsk array data.

There may be an additional uncertainty up to $40\%$ in $E_{em}$ assignment due to inadequate fitting of the longitudinal shower profile detected with HiRes. This may be the result of increased measurement error of the faint fluorescence light far from the shower maximum, and/or the direct and scattered Cherenkov light contamination to the signal.

There is an alternative $E_{em}$ estimation method proposed in~\cite{Muniz}. It is based on the relation between the number of electrons at the shower maximum, $N_{max}$, and energy of gamma-quanta initiating electromagnetic sub-cascade: $N_{max}=0.3E_\gamma/\sqrt{\ln E_\gamma}$~\cite{Belenky}. An advantage of the method is in avoidance of the cascade curve approximation; disadvantage consists in the asymmetric distribution of $N_{max}$. On the whole, it seems to be the more reliable approach to the estimation of electromagnetic component energy in comparison with integration of the longitudinal shower profile, not only at energies above $10^{19}$ eV, as was concluded in~\cite{Muniz}.

Presumably, the overall experimental uncertainty in ionization integral estimation is $\sim30\%$ for the Yakutsk array data, and $\leq45\%$ for the HiRes measurements. Consequently, notwithstanding nearly the same conversion factor from $E_{em}$ to $E_0$, the primary energy estimate may be distinctly different in these experiments due to systematic errors in $E_{em}$ evaluation.

\begin{figure}[t]
\center{\includegraphics[width=0.6\columnwidth]{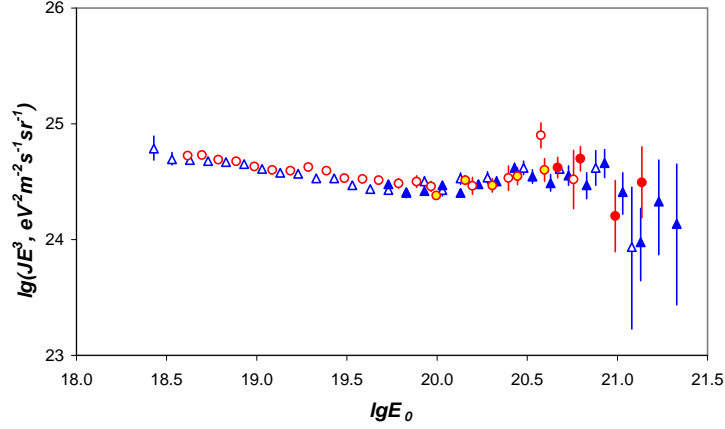}}
\caption{\footnotesize The energy spectrum of cosmic rays. Monocular data from two HiRes Eyes are given: HiRes-I (filled triangles) and HiRes-II (open triangles). The Yakutsk array data are selected with trigger-500 (open circles), and trigger-1000 (filled circles). Energy correction factors are given in the text.}
\label{fig:Spectra}
\end{figure}

To provide an illustration, the differential energy spectra measured with the Yakutsk array~\cite{CRIS} and HiRes~\cite{Sokolsky} are shown in Fig.~\ref{fig:Spectra}. The primary energy is decreased by $20\%$ for the former data, and increased by $30\%$ for the latter. Intensities are multiplied by $E_0^3\epsilon_0^3$ in order to demonstrate the spectrum features - 'ankle' around $10^{19}$ eV and steepening at $E_0\epsilon_0\sim10^{20}$ eV (Greisen-Zatsepin-Kuzmin effect). The shape and intensity of the spectra derived from the data of the Yakutsk array and HiRes are consistent indicating a need for systematic correction (within experimental errors) to the ionization integral measured.

\section{Conclusion}
Extremely different hadronic interaction models lead to the resultant (electromagnetic component energy/EAS primary particle energy) ratio consistent within 7\% if the model and primary mass guarantee the same maximum position in atmosphere and muon content of the shower:
$$
\frac{E_{em}}{E_0}=(0.87\pm0.01)+(0.02\pm0.01)\lg\frac{E_0\epsilon_0}{10^{18}},
$$
$10^{17}<E_0\epsilon_0<10^{19} eV$.

The primary energy estimation algorithms based on ionization integral measurement rely on this ratio and the resultant uncertainty originates from experimental errors predominantly; model dependent one is minor in the presence of contemporary measurements of the longitudinal shower development parameters.

\section*{Acknowledgment}
This work is partially supported by RFBR (grant \#06-02-16973) and  MSE (grant \#7514.2006.2).

\end{document}